\documentclass[%
superscriptaddress,
notitlepage,
twocolumn,
groupedaddress,
amsmath,amssymb,
prl,
floatfix,
]{revtex4-2}

\usepackage{graphicx}% Include figure files
\usepackage{bm}% bold math

\usepackage[x11names, rgb, html]{xcolor}

\definecolor{red}{HTML}{f54b1a}
\definecolor{pink}{HTML}{d19eb1}
\definecolor{orange}{HTML}{d3772e}
\definecolor{yellow}{HTML}{ebe85d}
\definecolor{green}{HTML}{0f6852}
\definecolor{lightblue}{HTML}{01abe9}
\definecolor{darkblue}{HTML}{1b346c}
\definecolor{tan}{HTML}{e5c39e}
\definecolor{darktan}{HTML}{af9e73}
\definecolor{grey}{HTML}{c3ced0}
\definecolor{darkgrey}{HTML}{9dadc4}
\definecolor{black}{HTML}{110d1b}
\definecolor{white}{HTML}{f1f8f1}

\usepackage{hyperref}
\hypersetup{
  colorlinks = true,
  linkcolor = red,
  citecolor = darkblue,
  urlcolor = lightblue
}
\usepackage{physics}

\def\lambdab{\boldsymbol{\lambda}}

\def\Thetab{\boldsymbol{\Theta}}

\def\etab{\boldsymbol{\eta}}

\def\vb{\boldsymbol{v}}

\def\xb{\boldsymbol{x}}

\def\Wb{\boldsymbol{W}}
\def\Xb{\boldsymbol{X}}

\def\grad{\nabla}

\def\RR{\mathbb{R}}

\newcommand{\avg}[1]{\left\langle #1 \right\rangle}

\def\<{\langle} \def\>{\rangle}

\DeclareRobustCommand{\argmin}{\operatorname*{argmin}}

\usepackage[normalem]{ulem}
\usepackage{algorithm}
\usepackage{algpseudocode}

\usepackage{amsmath}
\DeclareMathOperator\arctanh{arctanh}

\usepackage[english]{babel}

\makeatletter
\def\bbl@set@language#1{%
  \edef\languagename{%
    \ifnum\escapechar=\expandafter`\string#1\@empty
    \else\string#1\@empty\fi}%
  \@ifundefined{babel@language@alias@\languagename}{}{%
    \edef\languagename{\@nameuse{babel@language@alias@\languagename}}%
  }%
  \select@language{\languagename}%
  \expandafter\ifx\csname date\languagename\endcsname\relax\else
    \if@filesw
      \protected@write\@auxout{}{\string\select@language{\languagename}}%
      \bbl@for\bbl@tempa\BabelContentsFiles{%
        \addtocontents{\bbl@tempa}{\xstring\select@language{\languagename}}}%
      \bbl@usehooks{write}{}%
    \fi
  \fi}
\newcommand{\DeclareLanguageAlias}[2]{%
  \global\@namedef{babel@language@alias@#1}{#2}%
}
\makeatother

\DeclareLanguageAlias{en}{english}

\graphicspath{{figs/}}

\begin{document}

\title{A unified, geometric framework for nonequilibrium protocol optimization}

\author{Shriram Chennakesavalu}
\email{shriramc@stanford.edu}
\affiliation{Department of Chemistry, Stanford University, Stanford, CA 94305}%
\author{Grant M. Rotskoff}%
 \email{rotskoff@stanford.edu}
\affiliation{Department of Chemistry, Stanford University, Stanford, CA 94305}%

\date{\today}

\begin{abstract}
Controlling thermodynamic cycles to minimize the dissipated heat is a longstanding goal in thermodynamics, and more recently, a central challenge in stochastic thermodynamics for nanoscale systems. 
Here, we introduce a theoretical and computational framework for optimizing nonequilibrium control protocols that can transform a system between two distributions in a minimally dissipative fashion. 
These protocols optimally transport a system along paths through the space of probability distributions that minimize the dissipative cost of a transformation. 
Furthermore, we show that the thermodynamic metric---determined via a linear response approach---can be directly derived from the same objective function that is optimized in the optimal transport problem, thus providing a unified perspective on thermodynamic geometries.
We investigate this unified geometric framework in two model systems and observe that our procedure for optimizing control protocols is robust beyond linear response.
\end{abstract}

\maketitle

Understanding how to efficiently control thermodynamic cycles is a truly foundational problem in thermodynamics. 
Our modern mathematical framework for macroscopic thermodynamics emerged from efforts to describe the transfer of heat into work and to quantify the wasted or excess heat dissipated to the environment~\cite{carnot_reflexions_1872}.
As it has become possible to probe the thermodynamics of nanoscale systems, both experimentally and in computer simulations, the role of thermal fluctuations has reoriented our interpretation of the fundamental constraints imposed by the second law; at small scales, fluctuation theorems precisely quantify the relationship between entropy production and irreversible dynamics~\cite{crooks_nonequilibrium_1998,lebowitz_gallavotti_1999,kurchan_fluctuation_1998}. 
Exploiting nonequilibrium dynamics to understand equilibrium properties like free energy differences has been realized both experimentally and computationally via the Jarzynski equality~\cite{jarzynski_nonequilibrium_1997,liphardt_equilibrium_2002}.
However, the statistical accuracy of such calculations requires minimizing dissipation over the nonequilibrium ensemble of trajectories by choosing an appropriate external driving protocol~\cite{maragakis_bayesian_2008,rotskoff_geometric_2017}.
What is more, fundamental questions about the design and properties of nanoscale machines from biology to engineering require theoretical tools to carefully interrogate dissipation in stochastic systems. 

Despite the importance of measuring dissipation, it has proved challenging to do so accurately in nanoscale systems, with only indirect proxies available.
While bounds like the thermodynamic uncertainty relations~\cite{barato_thermodynamic_2015,gingrich_dissipation_2016} can be used to aid inference, these relations do not necessarily tightly constrain the dissipation and cannot be directly correlated with it in general~\cite{gingrich_inferring_2017}.
Nevertheless, recent experimental and computational advances have reinvigorated efforts to design optimal controllers for nanoscale systems.  
Optimizing a protocol through the use of ``thermodynamic geometry''~\cite{crooks_measuring_2007,salamon_thermodynamic_1983,salamon_length_1985,ruppeiner_thermodynamics_1979, schlogl_thermodynamic_1985}, an approach in which the dissipation is quantified through a Riemannian path length in the space of protocols, has proved among the most productive strategies for this problem~\cite{sivak_thermodynamic_2012,rotskoff_optimal_2015,rotskoff_geometric_2017,brandner_thermodynamic_2020}.
The metric itself is derived via a perturbative expansion~\cite{rotskoff_geometric_2017} and hence applies only in the limit of driving that is sufficiently slow or when the magnitude of the perturbation is sufficiently small.

Separately, initially spurred by developments in the study of variational solutions to certain partial differential equations~\cite{villani_topics_2003,jordan_variational_1998,benamou_monge_2002}, a distinct geometry, based on optimal transport theory, has been connected to nonequilibrium dissipation. 
In this formulation, distances are measured not with a Riemannian metric, but directly between probability distributions by determining a minimum cost transport plan that moves the probability mass from an initial distribution $\rho_A$ to a given target $\rho_B$. 
The cost defines the Wasserstein metric, which, in the Monge formulation, is formally defined through an optimization problem
\begin{equation}
    \mathcal{W}_2^2(\rho_A, \rho_B) = \inf_{T} \int_{\Omega} | \xb - T(\xb) |^2 \rho_A(\xb) d\xb,
    \label{eq:monge}
\end{equation}
where $T$ ranges over all valid maps or transportation plans that send $\rho_A$ to $\rho_B.$ 
The Wasserstein metric is a lower bound on the dissipative cost to transform $\rho_A$ to $\rho_B$ in a finite time, and, importantly, provides an alternate geometric framework for minimizing dissipation~\cite{aurell_refined_2012}.
Unlike the perturbative formulation that leads to the thermodynamic Riemannian metric, this approach makes no approximation to quantify the total change entropy along a dissipative transformation.
However, the constrained minimization problem~\eqref{eq:monge} that one must solve to compute the Wasserstein distance is notoriously challenging, both analytically and numerically~\cite{cuturi_sinkhorn_2013}.

Here, we introduce a theoretical and computational framework for optimizing control protocols that realize geodesics in the Wasserstein metric, which we refer to as displacement interpolations~\cite{mccann_convexity_1997}.
These are paths through the space of probability distributions that minimize the dissipative cost of the finite-time transformation. 
Furthermore, we show that the thermodynamic metric can be derived directly from the same objective function that we optimize in the optimal transport problem, emphasizing that the two coincide in the limit of slow driving.
Our result provides a unified geometric framework for minimizing the dissipative costs of nonequilibrium transformations. 
Crucially, the approach we propose is numerically tractable without globally computing the thermodynamic metric, a numerically costly procedure, especially when the dimensionality of the protocol is large.
We investigate this approach to minimum dissipation control in two simple models of nanoscale engines, and we compare control protocols determined via both the thermodynamic metric and the geometry of optimal transport. 
Remarkably, we observe that our procedure for optimizing control protocols is robust outside the linear response regime, which leads to a significant improvement in protocol design over the thermodynamic metric when the driving is fast. 

\paragraph{Connection between the thermodynamic metric and the optimal transport problem.}
We consider the problem of transforming an initial equilibrium distribution $\rho_A = \rho(\cdot, 0)$ into a target distribution $\rho_B = \rho(\cdot, t_{\rm f})$ with a nonequilibrium driving protocol $\lambdab$ of duration $t_{\rm f}.$
At sufficiently long times, we assume that the system relaxes to an equilibrium distribution $\rho_0(\xb) = e^{-\beta U(\xb, \lambdab)}/Z(\lambdab)$, where $U$ denotes the potential energy of the system. 
The results we derive below apply to systems that evolve according to overdamped Langevin dynamics, though we see that the methods we develop also apply to open quantum systems~\cite{vanvu_thermodynamic_2022}. While we do not consider underdamped Langevin dynamics here, minimizing dissipation in these systems is an active area of inquiry \cite{dago_dynamics_2022,muratore-ginanneschi_extremals_2014}.

We consider first a system with coordinates $\xb \in \Omega\subset \mathbb{R}^d$ subject to the following overdamped Langevin equation, 
\begin{equation}
    \dot{\xb} = -\nabla U\bigl(\xb, \lambdab(t)\bigr) + \sqrt{2\beta^{-1}} \etab(t)
    \label{eq:langevin}
\end{equation}
where $\etab$ is a Gaussian random variable with $\avg{\etab(t)}=0$ and $\avg{\etab_i(t)\etab_j(t')} = \delta(t-t') \delta_{ij}$.
The external protocol $\lambdab$ changes as a function of time, which drives the system away from equilibrium.
As a result, the time-dependent distribution $\rho(\xb,t)$ of states may be non-Boltzmann, but it does satisfy a Fokker-Planck equation
\begin{equation}
    \partial_t \rho + \nabla \cdot (\vb \rho) = 0,
    \label{eq:fp}
\end{equation}
where the local velocity is 
\begin{equation}
\vb(\xb,t) = -\nabla U\bigl(\xb(t),\lambdab(t)\bigr) - \beta^{-1} \nabla \log \rho(\xb,t).
\label{eq:localv}
\end{equation}
Non-conservative forces can be incorporated into this framework and only change the expression for the local velocity~\cite{aurell_refined_2012}.
% The instantaneous entropy of the system can be specified exactly through the Gibbs equation,
% \begin{equation}
%     \Sigma_{\rm sys}(t) = -k_{\rm B} \int_{\Omega} \rho(\xb,t) \log \rho(\xb,t) \ d\xb.
% \end{equation}
The total entropy production can be written in terms of the heat flow to the environment using the stochastic thermodynamics convention for the heat flow~\cite{sekimoto_langevin_1998,seifert_entropy_2005}
\begin{equation}
    \beta \mathcal{Q} = -\int \beta(t) \grad U\bigl(\xb(t), \lambdab(t)\bigr) \circ d\xb(t),
\end{equation}
and the Gibbs entropy of the system. 
Combining these terms, we obtain a quadratic form for the total entropy production along a nonequilibrium transformation~\cite{aurell_optimal_2011, aurell_refined_2012, supplement},
\begin{equation}
    \Delta \Sigma_{\rm tot} = \int_{0}^{t_{\rm f}} \beta(t) \int_{\Omega} \vb^T(\xb,t) \vb(\xb, t) \rho(\xb,t)\ d\xb\ dt.
    \label{eq:stot}
\end{equation}
Hence, to minimize the dissipation associated with a transformation from the thermodynamic state specified by $\lambdab(0)$ to a state in which the control parameters are fixed at $\lambdab(t_{\rm f}),$ we must identify a minimizer $\lambdab_*(t)$ of~\eqref{eq:stot}, noting that local velocity and the density both depend on $\lambdab$.

Formally, we solve
\begin{equation}
  \lambdab_* = \argmin_{\lambdab:[0,t_{\rm f}]\to \RR^k \quad \rho(\cdot,0) = \rho_A,\quad \rho(\cdot,t_{\rm f}) = \rho_B} \Delta \Sigma_{\rm tot}[\lambdab].
  \label{eq:bb1}
\end{equation}

This minimization problem has a geometric interpretation: the minimum of~\eqref{eq:stot} over all $(\vb,\rho)$ satisfying~\eqref{eq:fp} and~\eqref{eq:localv} with the boundary conditions that $\rho(\cdot,0)=\rho_A$ and $\rho(\cdot,t_{\rm f}) = \rho_B$ is exactly the Benamou-Brenier formulation of the Wasserstein optimal transport distance~\cite{benamou_monge_2002,peyre_computational_2019}.
In the present work, we optimize not over velocity fields and densities, but rather protocol $\lambdab$ which, when sufficiently flexible, provides substantial control over the distribution. While this constraint means that we may not saturate the optimal transport distance in general, the examples we consider here are ones in which the protocol provides complete control over the distribution. Furthermore, we believe that a protocol optimization framework is more physically practical than a setting in which the requisite velocity fields to minimize \eqref{eq:stot} are inaccessible to any external controller. 
Minimizing~\eqref{eq:stot} provides an alternative formulation of the Wasserstein optimal transport problem defined in~\eqref{eq:monge}, as explained in the supporting information~\cite{supplement}.
The connection between optimal transport and dissipation is well-known in the partial differential equations literature~\cite{villani_optimal_2009} and was subsequently connected to stochastic thermodynamics by Aurell~\cite{aurell_optimal_2011,aurell_refined_2012}; recently this connection was discussed in~\cite{nakazato_geometrical_2021} to provide a framework for general thermodynamic speed-limits (cf. \cite{aurell_refined_2012,salamon_length_1985,crooks_measuring_2007}).

Computing the minimizer $\lambdab_*$ analytically is generically challenging because of the nonlinear dependence of the distribution on the protocol.
Within a linear response approximation, the form of the distribution simplifies considerably, and the quadratic functional can be written as an explicit function of $\lambdab$; we now demonstrate that carrying out the minimization of $\eqref{eq:stot}$ with respect to $\lambdab$ recovers the Riemannian thermodynamic metric. 
Following the dynamical linear response framework of Zwanzig~\cite{zwanzig_nonequilibrium_2001, pavliotis_stochastic_2014}, we assume that $\lambdab$ changes slowly relative to the rate of relaxation of the system. 
We expand the instantaneous density around the equilibrium density with the control parameters fixed,
\begin{equation}
\label{eq:perturb_expansion}
    \rho(\xb,t) = \rho_0\bigl(\xb, \lambdab(t)\bigr) + \epsilon \rho_1\bigl( \xb, \lambdab(t) \bigr) + \mathcal{O}(\epsilon^2) + \dots
\end{equation}
At first order in $\epsilon$ we obtain
\begin{equation}
    \rho_1(\xb,t) = \beta \rho_0(\xb, t) \int_0^\infty \delta \Thetab(\xb^{\lambdab}(s),t) \cdot \dot \lambdab(t) ds
\end{equation}
where $\delta \Thetab(\xb,s)$ denotes the deviation of the generalized forces from their average, $\partial_{\lambdab} U (\xb,\lambda(t))-\partial_{\lambdab} F(\lambdab(t))$.
The only part of the quadratic functional that depends on $\lambdab$ is the work; to leading order in $\epsilon$ an explicit computation~\cite{supplement} shows that the protocol dependent dissipation is
\begin{equation}
    L[\lambdab] = \int_0^{t_{\rm f}} \dot{\lambdab}^T(t) \zeta\bigl( \lambdab(t) \bigr) \dot{\lambdab}(t) dt
    \label{eq:lrlen}
\end{equation}
with 
\begin{equation}
    \zeta \bigl(\lambdab(t)\bigr) = \beta \int_{0}^\infty \avg{\delta \Thetab(s) \delta\Thetab^T(0)}_{\lambdab(t)}  ds.
\end{equation}
Minimizing the expression~\eqref{eq:lrlen} yields a geodesic with respect to the positive definite symmetric form $\zeta$; these geodesics in protocol space are minimum dissipation protocols within the linear response approximation~\cite{crooks_measuring_2007,sivak_thermodynamic_2012}.
The $\mathcal{W}_2$-optimal local velocity associated with the thermodynamic geodesic is explicitly given by the negative, temperature scaled spatial gradient of the perturbative correction $\epsilon \rho_1,$ simply plugging \eqref{eq:perturb_expansion} into \eqref{eq:localv}.

\begin{figure}[ht]
    \centering
    \includegraphics[width=\linewidth]{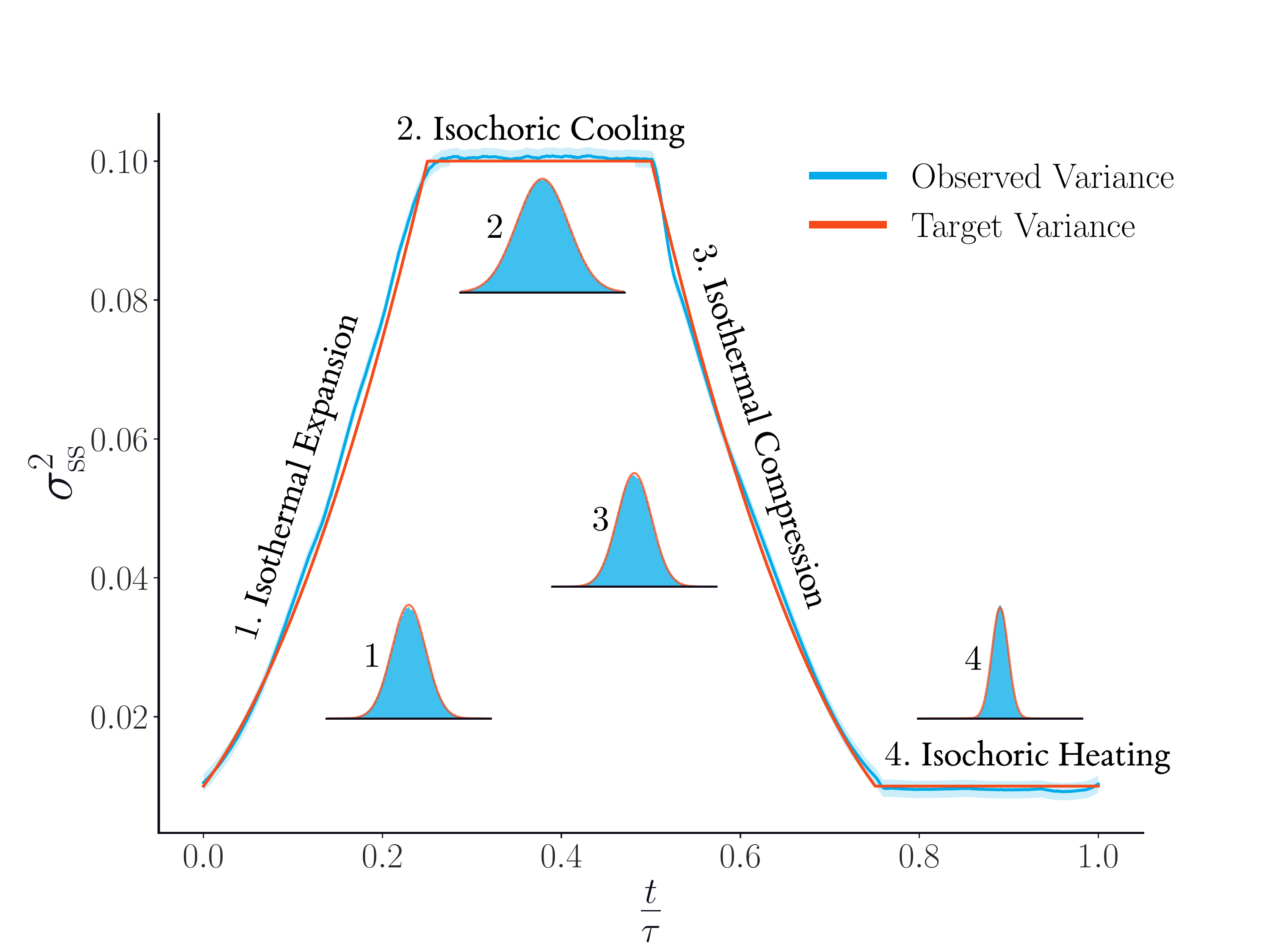}
    \caption{Brownian engine driven according to a Stirling cycle through target intermediate Gaussian distributions with mean $\mu=0$ and variance $\sigma^2_{\rm ss}$ .
    Variance of the target distributions on the Wasserstein geodesics (in red) along each stage are computed exactly between corresponding endpoints. In the fast-driving regime ($\tau = 10$), observed variance of trajectories under $\lambdab_*$ closely approximate target variance.}
    \label{fig:schematic}
\end{figure}

\paragraph{Computational approach.}
We consider two general paradigms for determining protocols that minimize \eqref{eq:stot}. 
The first involves specifying a set of intermediate distributions and learning a protocol $\lambdab_*(t)$ that drives the system along a Wasserstein geodesic between these intermediate distributions. 
Alternatively, we specify a protocol $\lambdab(t)$ and determine an optimal speed function $\phi_*(t)$ such that the system driven under $\lambdab(\phi_*(t))$ minimizes \eqref{eq:stot}. 
The first paradigm ensures that an engine reaches the desired intermediate distribution even when the driving is fast relative to the relaxation time of the system. 
The latter approach constrains the protocol, but is less computationally and experimentally demanding. 

To drive the system through prespecified intermediate distributions (see Fig.~\ref{fig:schematic}), we use gradient-based optimization to learn protocols $\lambdab_*(t)$. Given an initial distribution $\rho_A$ and a final distribution $\rho_B$, we compute a Wasserstein geodesic that interpolates these two distributions.
This path through the space of probability distributions is known as a displacement interpolation~\cite{villani_optimal_2009,peyre_computational_2019}. 
Computing this geodesic exactly is computationally demanding for arbitrary $\rho_A$ and $\rho_B$, so, for a general system, one must instead estimate a displacement interpolation using computational optimal transport algorithms, such as the Sinkhorn algorithm \cite{cuturi_sinkhorn_2013}.
We anticipate that this approximate calculation of the Wasserstein geodesic will be tractable even for relatively high-dimensional systems~\cite{altschuler_near-linear_2017}.
In the present work, we do not directly contend with this approximation; for the two minimal models considered here, we can compute the Wasserstein geodesics exactly.

Determining a displacement interpolation $\rho_*(t)$ does not directly yield a protocol that drives the system through the set of probability distributions that constitute the geodesic. 
While it would be possible to identify $\lambdab_*$ from $\rho_*$ in the quasi-static limit, when the system is driven far from equilibrium, we must optimize the protocol so that the empirical distribution remains close to $\rho_*(t)$. 
We represent $\lambdab_*$ using a neural network and carry out our optimization using automatic differentiation.
The gradient information is stored throughout the dynamical evolution of the system, and gradients are explicitly back-propagated through the trajectory.
Automatic differentiation is a natural way to optimize $\lambdab_*$ for the nonequilibrium systems considered here, as it enables dynamical information to be implicitly incorporated into a learned protocol. 
We note that for large protocol durations, it is impractical to differentiate throughout the entire trajectory; instead, we carry out our optimization over short-time intervals.

\begin{figure}[ht]
\includegraphics[width=0.95\linewidth]{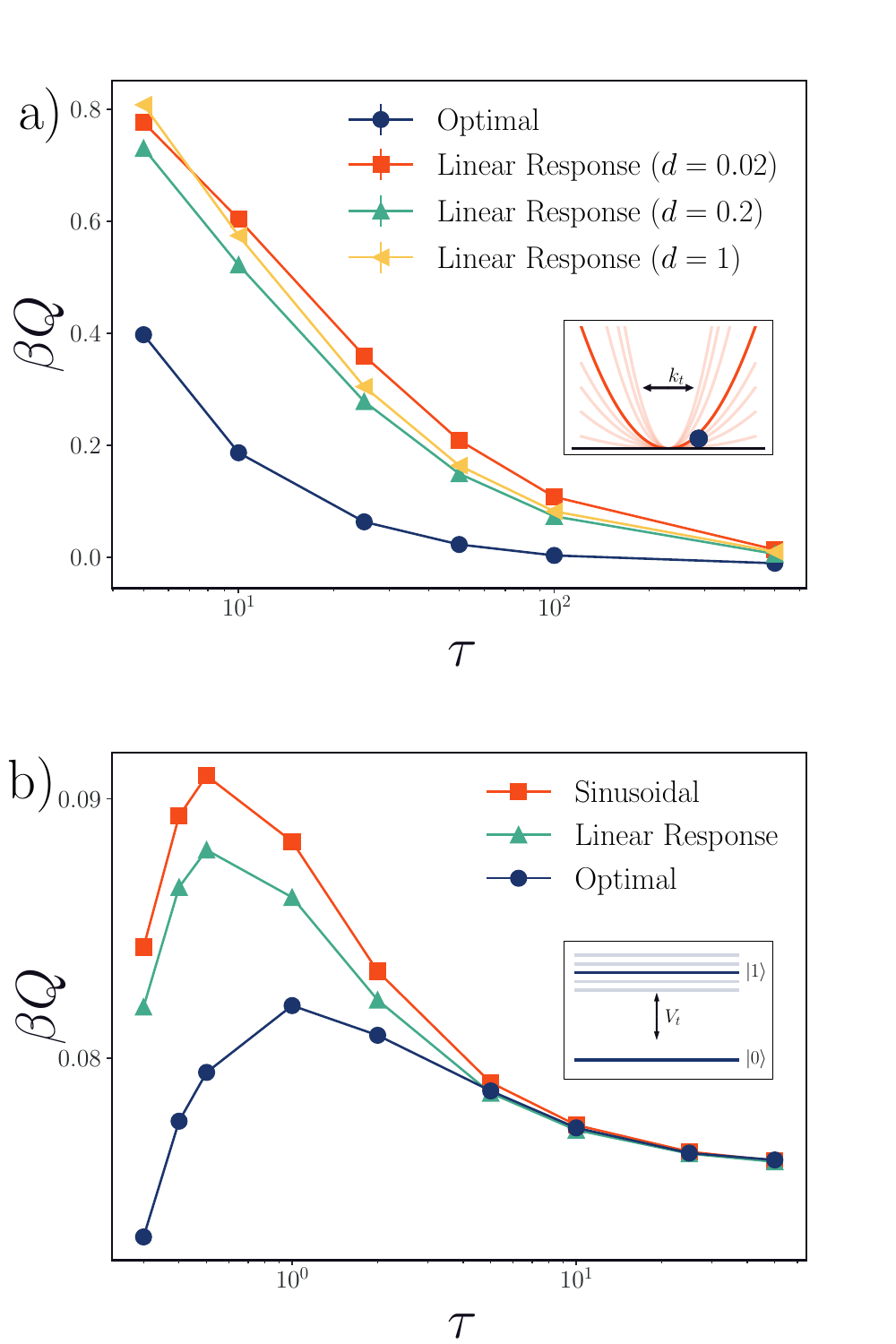}
\caption{ The dissipation profiles for two nanoscale engines plotted across different protocol durations $\tau$. a) For a nanoscale Brownian engine (see inset), the dissipation for the optimal protocol $\lambdab_*$ (dark blue circle) is lower than the dissipation for protocols computed according to a linear response approximation $\lambdab_{\rm LR}$ for different smoothness parameters $d$ \cite{brandner_thermodynamics_2015}. b) For a superconducting qubit (see inset), the optimal protocol $\lambdab(\phi_*)$ (dark blue circle) is less dissipative than a protocol computed using a linear response approximation $\lambdab(\phi_{\rm{LR}})$ (green triangle) \cite{brandner_thermodynamic_2020} or a base sinusoidal protocol $\lambdab(t)$ (red square)}
\label{fig:models}
\end{figure}

\paragraph{Optimizing a nanoscale Brownian engine.}
We consider a minimal model for a nanoscale Brownian engine,
consisting of a Brownian particle in a harmonic potential.
The temperature $T$ of the heat bath and the strength of the harmonic potential $k$ are controlled to mimic a Stirling cycle~\cite{blickle_realization_2012}. 
The model is depicted schematically in the inset of Fig.~\ref{fig:models} (a).
In our implementation, each stage of the engine cycle has a fixed duration of $\tau/4$.
Given $T_h$ and $T_c$, the maximum and minimum allowed temperature of the bath, and $k_h$ and $k_l$, the maximum and minimum allowed strength of the harmonic potential, we can exactly specify the equilibrium distributions at the endpoints of each step of the Stirling cycle, which are Gaussian with $\mu = 0$ and $\sigma^2 = T/k$.
Because the distributions are Gaussian, we can exactly determine a displacement interpolation $\rho_*$ between the endpoints of each stage.

\begin{figure}[ht]
\includegraphics[width=0.85\linewidth]{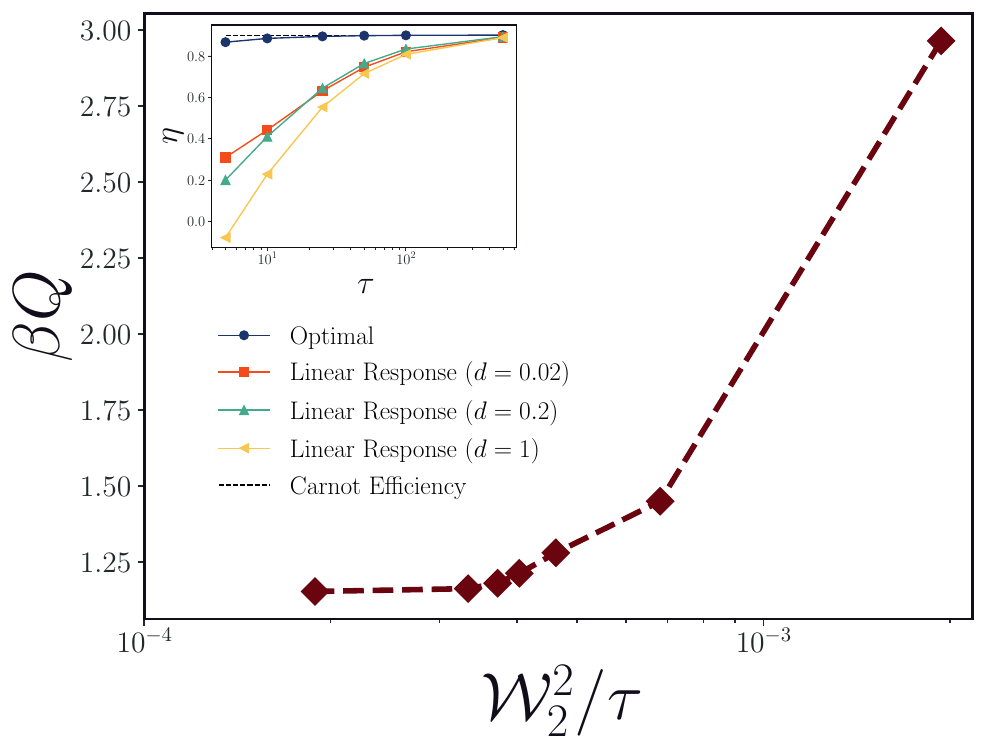}% Here is how to import EPS art
\caption{The dissipation plotted against $\mathcal{W}_2^2/{\tau}$ (the squared Wasserstein distance normalized by protocol duration) for the isothermal compression of a Brownian engine across different $\tau$. An increasing dissipative cost is incurred as $\mathcal{W}_2^2/{\tau}$ increases, consistent with \eqref{eq:stot}. In the inset, the efficiencies of the Brownian engine plotted against the protocol duration $\tau$ for a learned optimal protocol $\lambdab_*$ and protocols determined from a linear approximation \cite{brandner_thermodynamics_2015}. Even for fast-drivings, the Brownian engine driven by $\lambdab_*$ (dark blue circle) can achieve near Carnot efficiency and has a higher efficiency in comparison to the Brownian engine driven by $\lambdab_{\rm LR}$.}
\label{fig:w2vL}
\end{figure}

We carry out this optimization for a range of protocol durations $\tau$, the shortest of which are far from the linear response regime and the longest of which are essentially quasi-static.
For each $\tau$ we consider, we optimize the protocol $\lambdab_*$ separately.
We compare our results with optimal protocols determined using the linear response approximation~\cite{brandner_thermodynamics_2015}, which we denote $\lambdab_{\rm LR}$. 
Here, $\lambdab_{\rm LR}$ also drives the engine along a Stirling cycle and has the same $T_h$, $T_c$, $k_h$ and $k_l$ as $\lambdab_*$. Importantly, we observe that, in the fast-driving regime, the Brownian engine driven under $\lambdab_*$ is significantly less dissipative than the Brownian engine driven under $\lambdab_{\rm LR}$, as shown in Fig.~\ref{fig:models}. 
As the engine undergoes isothermal compression---the most dissipative step---the Brownian engine under $\lambdab_{\rm LR}$ significantly deviates from the displacement interpolation. However, under $\lambdab_{*}$, the Brownian engine is able to closely realize this geodesic, resulting in a consistently lower dissipation profile. 
Finally, for $\lambdab_{*}$ we further analyze the relationship between $\mathcal{W}_2^2/\tau$ and the dissipation for this step and observe that as $\mathcal{W}_2^2/\tau$ increases, the dissipation increases (Fig.~\ref{fig:w2vL}), as expected from~\eqref{eq:stot}.

Remarkably, we see that the Brownian engine driven under $\lambdab_*$ enables us to realize protocols that have a higher efficiency $\eta$ compared to protocols driven by the linear response protocol $\lambdab_{\rm LR}$ (see Fig~\ref{fig:w2vL} inset). 
In fact, the optimal transport protocols approach the Carnot efficiency even for fast driving. 
In this regime, the linear response protocol $\lambdab_{\rm LR}$ does not maximally expand the engine, thus limiting the work done by it. 
Because we learn a distinct $\lambdab_{*}$ for each $\tau$, the Brownian engine \emph{does} expand nearly maximally, resulting in greater work extraction. 
Crucially, the system also remains near the displacement interpolants, ensuring minimal dissipation and thus a higher efficiency.

\paragraph{Minimum dissipation control of a model superconducting qubit}

To show that our computational framework also applies to Markovian quantum dynamics, we investigate a model of a superconducting qubit engine.
We note that for discrete state Markov processes, the extending the Benamou-Brenier formulation exactly requires a distinct metric~\cite{maas_gradient_2011}, however, we optimize $\mathcal{W}_1$ and demonstrate numerically that this lower bound to heat (cf. \cite{vanvu_thermodynamic_2022} Thm. 2) also yields a powerful variational principle for optimal control problems.
In this model (Fig.~\ref{fig:models} (b) inset), we modulate the temperature $T$ of the environment and the level-splitting of the qubit $V$ as a function of time. 
We optimize protocols i) using a pre-specified protocol and varying the speed at which the protocol is traversed and, ii) by setting target intermediate distributions and optimizing $V(t)$ and $T(t)$.

For (i), we used a sinusoidal protocol $\lambdab(t)$, as specified in the supplementary material~\cite{supplement}.
We compare three different protocols $\lambdab(t)$, $\lambdab(\phi_{\rm LR}(t))$ and
$\lambdab(\phi_{*}(t))$, where $\phi_{*}$ was determined by minimizing the length $\mathcal{W}_2$ along the protocol and $\phi_{\rm LR}$ was determined in \cite{brandner_thermodynamic_2020}. 
We observe that for the qubit engine driven under $\lambdab(\phi_{*}(t))$, the dissipation is lower compared to the engine driven by both  $\lambdab(t)$ and $\lambdab(\phi_{\rm LR}(t))$ for fast-driving, as shown in Fig. 2 (b). 
We find that the work obtained using $\lambdab(\phi_{\rm LR}(t))$ is higher than that of $\lambdab(\phi_{*}(t))$, resulting in the efficiency under $\lambdab(\phi_{\rm LR}(t))$ being marginally better than that of $\lambdab(\phi_{*}(t))$.
Of course, the expression that we optimize~\eqref{eq:stot} only includes dissipation as an objective, so there is no guarantee of higher efficiency.

For (ii), we optimize a protocol $\lambdab_*$ to drive the qubit engine through prespecified intermediate distributions $\rho_*(t)$. We considered the distributions that the system relaxed to at $t = 0, \tau/4, \tau/2$ and $3\tau/4$ in the quasi-static limit under the sinusoidal protocol, and specified $\rho_*$ to be the displacement interpolation of these distributions. We computed these geodesics exactly using the density matrix. As with the Brownian heat engine, we used automatic differentiation to learn a protocol $\lambdab_*(t)$. The optimal protocol $\lambdab_{*}(t)$ is less dissipative than all other protocols, and the corresponding steady state distribution  $\hat{\rho}(t)$ closely realizes $\rho_*$. Ultimately, the results observed in the two systems considered demonstrate the utility of the unified geometric framework and the computational approach introduced here in learning minimally dissipative protocols for nonequilibrium control.

\bibliography{references}
\clearpage
\onecolumngrid
\setcounter{equation}{0}
\setcounter{figure}{0}
\renewcommand{\theequation}{S\arabic{equation}}
\renewcommand{\thefigure}{S\arabic{figure}}

\section{Derivation of the Thermodynamic Metric}

We consider a system with coordinates $\xb\in \RR^d$ subject to an ergodic Markovian dynamics.
A particle trajectory in which the $i$th particle is at $\xb^{(i)}_0$ at $t=0$ is denoted $\Xb^{(i)}(t; \xb_0^{(i)})$.
Within this framework, the density
\begin{equation}
  \rho(\xb, t) = \avg{\delta(\Xb(t;\xb_0)-\xb)}
  \label{eq:lrho}
\end{equation}
evolves according to a linear partial differential equation
\begin{equation}
  \partial_t \rho(\xb,t) = \mathcal{L}^{\dagger} \rho(\xb,t).
\end{equation}
For example, we consider the microscopic dynamics specified by the stochastic differential equation
\begin{equation}
  d\Xb(t) = -\nabla U(\Xb(t), \lambdab(t)) dt + \sqrt{2\beta^{-1}} d\Wb_t,
\end{equation}
where $\lambdab(t) \in \RR^k$ is a vector of control parameters conjugate to the generalized forces $\Thetab(\xb,t) = -\partial_{\lambdab} U(\xb, \lambdab(t))$, $\beta$ is the inverse temperature, and $U$ is the potential energy of system, and $\Wb$ is a standard Wiener process.
In this case, the distribution satisfies the evolution equation
\begin{equation}
  \partial_t \rho(\xb,t) = -\nabla \cdot (\vb(\xb,t) \rho(\xb,t))
  \label{eq:afp}
\end{equation}
where the local velocity $\vb$ is explicitly
\begin{equation}
  \vb(\xb, t) = -\nabla U(\xb, \lambdab(t)) - \beta^{-1} \nabla \log \rho(\xb,t)
\end{equation}
meaning that
\begin{equation}
  \mathcal{L}^\dagger = -\nabla U \cdot \nabla + \beta^{-1} \Delta.
\end{equation}

In this work, we focus on the problem of transforming an initial distribution, which could be the equilibrium distribution with the initial control parameters, e.g.,
\begin{equation}
  \rho_A(\xb) = Z^{-1}(\lambdab(0)) e^{-\beta U(\xb,\lambdab(0))},
\end{equation}
into a target distribution $\rho_B(\xb)$ which is the distribution that the system relaxes to when the control parameters are fixed at $\lambdab(t_{\rm f})$.
The protocol has a finite duration $0<t_{\rm f}< \infty$.
More general distributions $\rho_A$ and $\rho_B$ could also be considered.
We employ the standard stochastic thermodynamics definitions of work,
\begin{equation}
  W[\Xb(t)] = \int_0^{t_{\rm f}} \Thetab(\Xb(t,\xb_0)) \cdot \dot{\lambdab}(t) dt,
  \label{eq:work}
\end{equation}
and heat,
\begin{equation}
  \mathcal{Q}[\Xb(t)] = -\int_0^{t_{\rm f}} \nabla U(\Xb(t,\xb_0)) \circ d\Xb(t).
  \label{eq:heat}
\end{equation}
In this formulation, the work and the heat yield the total change in energy when the system is driven by the protocol, meaning that the first law of thermodynamics is simply,
\begin{equation}
    \Delta U = W - \mathcal{Q}.
\end{equation}
Using these expressions and \eqref{eq:afp}, the total entropy change is given by~\cite{aurell_refined_2012},
\begin{equation}
  \Delta \Sigma_{\rm tot} = \int_0^{t_{\rm f}} \beta(t) \int_{\Omega} \vb(\xb,t)^T \vb(\xb,t) \rho(\xb,t) d\xb\ dt
  \label{eq:astot}
\end{equation}
where the excess entropy dissipated to the environment is given by the average dissipated heat
\begin{equation}
    \Delta \Sigma = \beta(\avg{W-\Delta U}) = \beta \avg{\mathcal{Q}}.
\end{equation}
The average energy change of the system is a state function and has no explicit dependence on the protocol used.

To find a minimum dissipation protocol $\lambdab_*$, we must solve the following minimization problem:
\begin{equation}
  \lambdab_* = \argmin_{\lambdab:[0,t_{\rm f}]\to \RR^k} \Delta \Sigma_{\rm tot}[\lambdab] \textrm{ subj. to } \rho(\cdot,0) = \rho_A,\quad \rho(\cdot,t_{\rm f}) = \rho_B.
  \label{eq:bb}
\end{equation}
As discussed in the main text, this is the Benamou-Brenier formulation of the optimal transport problem~\cite{benamou_monge_2002}.
It should be noted that the rigorous formulation of the relation between the minimizer \eqref{eq:bb} and the optimal transportation plan requires that the time-dependent local velocity fields and density $(\vb, \rho)$ are allowed to vary over all possible functions that satisfy~\eqref{eq:fp} and the boundary conditions.
While we place no restrictions on the control parameters $\lambdab$, in practice we represent $\lambdab$ with a neural network, so we do restrict the function class to an extent, though sufficiently wide neural networks should allow us to find expressive enough protocols to ensure that this restriction is unimportant.

\textcite{benamou_computational_2000} initially established the connection between the Monge problem~\eqref{eq:monge} and formulation of the optimal transport problem as a minimization over velocity fields~\eqref{eq:stot}.
We outline the core argument here for completeness and because it is not widely known in the physics community. 
First, we write the dynamical evolution in the state space in the Lagrangian formalism, as above $\Xb(t; \xb)$ with the initial condition $\Xb(0; \xb) =\xb$.
The associated particle velocity we denote
\begin{equation}
    \dot{\Xb}(t; \xb) = v(\xb,t),
    \label{eq:lv}
\end{equation}
emphasizing that though we have changed the representation the conservation equation $\partial_t\rho + \nabla\cdot (\rho v) = 0$ still holds for \eqref{eq:lrho} and \eqref{eq:lv}. 
The boundary conditions of the optimal transport problem require that an admissible solution in terms of the density~\eqref{eq:lrho} and~\eqref{eq:lv} must have the property that at some fixed final time $t_{\rm f}$, for all bounded, continuous functions $f$,
\begin{equation}
    \int_{\Omega} f(\xb) \rho_B(\xb) d\xb = \int_{\Omega} f(\Xb(t_{\rm f}; \xb)) \rho_A(\xb) d\xb.  
\end{equation}
Obviously, if $\Xb(t_{\rm f}; \xb) \equiv T(\xb)$, the optimal map, is a valid solution.

They additionally show that
\begin{equation}
    t_{\rm f} \int_{\Omega} \int_0^{t_{\rm f}} |\dot{\Xb}(t,\xb)|^2 \rho_A(\xb)\ d\xb\ dt \geq \int_{\Omega} |{\Xb}(t,\xb)-\xb|^2 \rho_A(\xb)\ d\xb\ dt \geq \int_{\Omega} |T(\xb)-\xb|^2 \rho_A(\xb)\ d\xb\ dt.
\end{equation}
The first inequality is Jensen's inequality; the second is due to the fact that $T$ minimizes the integral.
Hence, the optimal map is simply given by
\begin{equation}
    \Xb(t; \xb) = (1-\frac{t}{t_{\rm f}}) \xb + \frac{t}{t_{\rm f}} T(\xb),
\end{equation}
which also defines the optimal pair $(\rho, v)$ via \eqref{eq:lrho} and \eqref{eq:lv}.

We do not know the distribution $\rho(\xb,t)$ at all intermediate times without solving~\eqref{eq:afp} because, as we change the protocol, the system is driven away from equilibrium.
However, for sufficiently weak driving, we can solve the equation for the dynamics perturbatively.
We first assume that the rate of driving is slow; explicitly, $\dot{\lambdab} \equiv \epsilon \dot{\lambdab}_\epsilon$ so we can write the adjoint linear operator driving the dynamics as
\begin{equation}
  \mathcal{L}^{\dagger} = \mathcal{L}_0^{\dagger} + \epsilon \mathcal{L}_1^{\dagger}
\end{equation}
with
\begin{equation}
  \mathcal{L}_1^\dagger = \dot{\lambdab}_\epsilon \cdot \partial_{\lambdab}.
\end{equation}
Expanding the solution to first order in $\epsilon$,
\begin{equation}
  \rho(\xb,t) = \rho_0(\xb,t) + \epsilon \rho_1(\xb,t) + \mathcal{O}(\epsilon^2),
\end{equation}
we see that
\begin{equation}
  \rho_0(\xb,t) = e^{-\beta U(\xb,\lambdab(t))}/Z(\lambdab(t))
\end{equation}
and the correction term satisfies
\begin{equation}
  \partial_t \rho_1(\xb,t) = \mathcal{L}_0^\dagger \rho_1(\xb,t) + \mathcal{L}_1^\dagger \rho_0(\xb,t)
\end{equation}
at order $\epsilon$.
Using the method of characteristics and the initial condition $\rho_1(\xb,0) = 0$ and assuming that the protocol is advanced from time $t$, we obtain
\begin{equation}
  \rho_1(\xb,t) = \int_0^t e^{\mathcal{L}_0^\dagger(\tau-s)} \dot{\lambdab}_\epsilon(t) \cdot \partial_{\lambdab(t)} \rho_0(\xb^{\lambdab}(s),t) ds.
\end{equation}
Here $\xb^{\lambdab}$ denotes the fast-timescale relaxation of the system to slow driving and $\xb^{\lambdab}(0) = \xb_0$; the protocol is effectively fixed over the relaxation time of $\xb^{\lambdab}$ due to the assumed timescale separation.
In the integral, $\tau$ denotes the relaxation of the fast process, which typically can be taken to infinity assuming that the integral converges sufficiently quickly.
Writing the equilibrium distribution as
\begin{equation}
  \rho_0(\xb,t) = e^{-\beta U(\xb,\lambdab(t)) + \beta F(\lambdab(t))},
\end{equation}
where $F$ is the equilibrium free energy with the control parameters fixed,
\begin{equation}
  F(\lambdab) = -\beta^{-1} \log \int_{\Omega} e^{-\beta U(\xb,\lambdab)} d\xb.
\end{equation}
The derivative with respect to $\lambdab$ yields the deviation of the generalized forces from their average value at a particular point $\xb$, that is, 
\begin{equation}
  \partial_{\lambdab(t)} \rho_0(\xb,t) = \beta\left( \Thetab(\xb(s),t) - \avg{\Thetab(\xb(0),t)}_{\lambdab(t)} \right) \rho_0(\xb,t).
\end{equation}
Because the protocol advances slowly relative to the timescale of relaxation of the system, we can extend $\tau\to \infty$ so that
\begin{equation}
  \epsilon \rho_1(\xb,t) = \epsilon \beta \rho_0(\xb, t) \int_0^\infty \delta \Thetab(\xb^{\lambdab}(s),t) \cdot \dot \lambdab(t) ds.
\end{equation}
The excess dissipation associated with the protocol can be computed by computing the average work associated with the perturbation.
%The contribution from the total change in energy does not depend on the protocol and contributes the leading term, so we are left with, at order $\epsilon$, 

This expression optimizes \eqref{eq:astot} and recovers the thermodynamic metric of~\citet{sivak_thermodynamic_2012}.
The dissipation along a protocol $\lambdab:[0,t_{\rm f}]$ is proportional to the path functional
\begin{equation}
    L[\lambdab]:= \int_0^{t_{\rm f}} \dot{\lambdab}^T(t) \zeta(\lambdab(t)) \dot{\lambdab}(t) dt,
\end{equation}
where 
\begin{equation}
    \zeta(\lambdab(t)) = \beta \int_0^\infty \avg{\delta \Thetab(\xb(s),t) \delta \Thetab^T(\xb(0),t)}_{\lambdab} ds;
\end{equation}
the subscript $\lambdab$ denotes that the average for the initial condition is taken from the equilibrium distribution with the control parameters fixed at $\lambdab(t)$.
Because $\zeta$ is a positive semi-definite form, it is a Riemannian metric and all associated path lengths are non-negative.

\section{Computational Details for  Protocol Optimization}
We consider two different approaches for learning protocols that minimize \eqref{eq:stot}: we either specify target intermediate distributions and learn a protocol $\lambdab_*$ that drives the system through these distributions or specify a protocol $\lambdab$ and determine $\phi_*$, the speed at which the protocol is traversed.

We use automatic differentiation to learn a time-dependent $\lambdab_*$ that drives the system through target intermediate distributions. Here, $\lambdab_*:\RR^1 \to \RR^2$ is a single hidden layer neural network. The input to $\lambdab_*$ is the current time of the protocol and the output is the temperature and the mechanical parameter being modulated. Our general algorithm is shown below in Algorithm~\ref{alg:autodiff}.

\begin{algorithm}[H]
    \caption{Automatic Differentiation to Optimize $\lambdab_*$} \label{alg:autodiff}
    \begin{algorithmic}
    \State {\bfseries Initialize protocol $\lambdab_*$}
    \For {$e = 1, \dots, n_{\rm{epochs}}$}
    \State Initialize state ${\xb_0}$
        \For {$t < \tau$} 
            \State{$\Xb^{\lambdab}(t+ \Delta t) = \Xb^{\lambdab}(t) + \dot{\Xb}^{\lambdab}(t) \Delta t$} \Comment{Use desired integration scheme}
            \State{$t \leftarrow t + \Delta t$}
            \If{end of interval}
                \State{Compute $\mathfrak{L}(t)$ and update $\lambdab_*$}
                \State{Clear gradient information}
            \EndIf
        \EndFor
    \EndFor
    \end{algorithmic}
\end{algorithm}

The automatic differentiation approach introduced above can be extended to any integration scheme that stores the exact gradients along the integration step. Alternatively, an adjoint method can be used, where gradients are computed by integrating backwards in time \cite{chen_neural_2018}, thus incurring a lower memory cost. However, for the systems we investigate here, it is tractable to store the gradients and optimize $\lambdab_*$ by directly backpropagating through the dynamics. 

When the total number of time steps is large for a longer protocol duration $\tau$, it is impractical to differentiate through the entire trajectory. During backpropagation, gradient information from each time step is multiplied, via the chain rule, leading to exploding or vanishing gradients \cite{metz_gradients_2021}. To mitigate this, we split our trajectory into $N$ short-time intervals and instead carry out intermediate optimization steps during a cycle. In practice, we see that an interval length, $M = \frac{\tau}{N dt}$, of around 5-10 steps is ideal, with larger $M$ for larger $\tau$ being more expedient. See \url{https://github.com/rotskoff-group/wasserstein-interpolation} for code and the parameters used.

%See \shriram{github} for the full list of parameters used.

\section{Nanoscale Brownian Engine}
We investigated a system consisting of a one-dimensional overdamped Brownian particle in a heat bath of temperature $T$ confined by a harmonic potential with time-dependent strength $k$,
\begin{equation}
    U_{\rm harmonic}(X(t), \lambdab(t)) = \frac{1}{2} k(t) X(t)^2,
    \label{eq:harmonic}
\end{equation}
with the equation of motion evolving according to an overdamped Langevin equation
\begin{equation}
    dX(t) = -\nabla U_{\rm harmonic}(X(t), \lambdab(t)) dt + \sqrt{2\beta^{-1}} dW_t,
    \label{eq:brownian_eom}
\end{equation}
where $W$ is standard Wiener process.

The Brownian engine is driven according to the Stirling cycle, where each stage is of equal duration $\tau/4$. We can fully characterize the cycle given $T_h$, $T_c$, $k_h$ and $k_l$, the maximum and minimum allowed temperature and trap strength respectively. 
Additionally, because the equilibrium distributions are Gaussian with $\mu = 0$ and $\sigma^2 = T/k$, we can exactly specify intermediate distributions along the cycle, because the displacement interpolation that minimizes the Wasserstein distance is Gaussian at each intermediate time in this case~\cite{peyre_computational_2019}. 
We consider the four Gaussian distributions that happen at the beginning of each stage, which have variance $\sigma^2_l$, $\sigma^2_h$, $\sigma^2_h$, and $\sigma^2_l$ respectively, where $\sigma^2_l$ = $T_h/k_h$ and $\sigma^2_h$ = $T_c/k_l$. Along each stage, we compute displacement interpolations $\rho_*(t)$ that interpolate the corresponding endpoints. This geodesic defines a set of intermediate distributions, and we learn a protocol $\lambdab_*$ that can successfully drive the engine so that the observed distribution $\hat{\rho}(t)$ is the same as the target distribution $\rho_*(t)$. Because the endpoint distributions are Gaussian, the distributions along the geodesic $\rho_*(t)$ are likewise Gaussian with $\mu_t = 0$ and variance
\begin{equation}
    \sigma_t^2 = \biggl(\frac{t_f - t}{t_f} \sigma_i + \frac{t}{t_f} \sigma_f\biggr)^2,
\end{equation}
where $\sigma^2_i$ is the variance of $\rho_*(0)$ and $\sigma^2_f$ is the variance of $\rho_*(t_f)$, with $0 \leq t \leq t_f$.

We compare the trained $\lambdab_*$ to $\lambdab_{\rm LR}$, a protocol determined using a linear response approximation \cite{brandner_thermodynamics_2015} that likewise drives the Brownian engine according to the Stirling cycle. To ensure that the protocols were comparable, the maximum and minimum allowed temperatures and harmonic potential strengths were the same for $\lambdab_{\rm LR}$ and $\lambdab_*$. We used an Euler-Maruyama integration scheme to simulate the dynamics of the Brownian engine and implemented this in Python using the \texttt{PyTorch} and \texttt{torchsde} packages. Because the distributions of this system are 
 Gaussian, we optimized a Mean Squared Error (MSE) loss 
 \begin{equation}
     \mathfrak{L} (t) = (\sigma^2_{\hat{\rho}(t)} - \sigma^2_{\rho_*(t)})^2 + (\mu_{\hat{\rho}(t)} - \mu_{\rho_*(t)})^2.
 \end{equation}
For more complex distributions, it will be expedient to use loss functions, such as the Kullback-Leibler (KL) divergence to estimate the instantaneous derivation of the empirical distribution from the target displacement interpolant. 
Here, $\hat{\rho}$, the empirical distribution of positions simulated under $\lambdab_*$, was determined by simulating 1000 independent Brownian engines in parallel.

% For the cycle, we define 4 parameters: $\sigma^2_h$, $\sigma^2_l$,
% $T_l$ and $T_h$ and aimed to learn a time-dependent protocol $\Lambda_t \equiv (T_t, k_t)$ that minimized the 
% $\mathcal{W}_2$ distance along the cycle. The first step of the cycle consists of an isothermal expansion at $T_h$, where 
% $k$ is decreased from $\frac{T_h}{\sigma_l^2}$ to $\frac{T_h}{\sigma_h^2}$. 
% The second step of the cycle consists of an isochoric cooling, where T is decreased from $T_h$ to $T_l$ and k
% is decreased from $\frac{T_h}{\sigma_h^2}$ to $\frac{T_l}{\sigma_h^2}$. The third step of the cycle consists
% of an isothermal compression at $T_l$, where $k$ is increased from $\frac{T_l}{\sigma_h^2}$ to $\frac{T_l}{\sigma_l^2}$. 
% The fourth and final step consists of an isochoric heating, where T is increased from $T_l$ to $T_h$ and
% $k$ is increased from $\frac{T_l}{\sigma_l^2}$ to $\frac{T_h}{\sigma_l^2}$. 

We compute the heat produced along a cycle as 
\begin{equation}
    Q =  \int -\partial_x U_{\rm harmonic}(X_t, \lambdab(t)) \circ dX_{t},
    \label{eq:brownian_heat}
\end{equation}
where $\circ$ is the time-symmetric Stratonovich product.
Similarly, the work done on the system along a trajectory is 
\begin{equation}
    W =  \int \partial_k U_{\rm harmonic}(X_t, \lambdab(t))\ dk.
    \label{eq:brownian_work}
\end{equation}
To account for the time varying temperature, we note that we compute the dissipation along the trajectory as  
\begin{equation}
    \beta Q =  \int \frac{-\partial_x U_{\rm harmonic}(X_t, \lambdab(t)) \circ dX_{t}}{k_{\rm B} T_t}.
    \label{eq:brownian_dissip}
\end{equation}
With these definitions, we define the efficiency $\eta$ of the engine as
\begin{equation}
    \eta =  \frac{\avg{W}}{\avg{Q_h}},
    \label{eq:efficiencyq}
\end{equation}
where 
\begin{equation}
    Q_h =  \int \gamma_t \bigl(-\partial_x U_{\rm harmonic}(X_t, \lambdab(t))\bigr) \circ dX_{t},
    \label{eq:brownian_heat_hot}
\end{equation}
and 
\begin{equation}
    \gamma_t = \frac{T_h(T_l - T_t)}{T_t(T_l - T_h)}
    \label{eq:brownian_gamma}
\end{equation}
as defined in \cite{brandner_thermodynamics_2015}. See Fig~\ref{fig:models} a) and Fig~\ref{fig:w2vL} inset for the dissipation and efficiency of the Brownian engine driven by these $\lambdab_*$ and $\lambdab_{\rm LR}$. Finally, we compute the squared Wasserstein distance $\mathcal{W}_2^2$ along the isothermal steps of the cycle
\begin{equation}
    \mathcal{W}_2^2 = \sum_0^{\tau/(4 dt)}\bigl((\sigma_{(i + 1)dt} - \sigma_{(i)dt})^2 + ((\mu_{(i + 1)dt} - \mu_{(i)dt})^2\bigr)
\end{equation}
and plot the relationship between the dissipation, $\beta Q$, and $\mathcal{W}_2^2/\tau$ for the isothermal compression step of the Stirling cycle in Fig.~\ref{fig:w2vL}.

\section{Qubit Engine}
We model a system of a single superconducting qubit evolving according to Lindbladian dynamics as introduced in \cite{brandner_thermodynamic_2020}. 
Throughout we work with probability distributions over the eigenstates $\ket{0}$ and $\ket{1}$.
This choice removes any ambiguity concerning the definition of a ``Quantum'' Wasserstein distance over density matrices~\cite{agredo_quantum_2017,depalma_quantum_2021,vanvu_thermodynamic_2022}, but does require an eigenbasis which could be challenging in general.
The Hamiltonian of this system is 
\begin{equation}
    H_v = -\frac{\hbar\Omega}{2}\bigl(\epsilon\sigma_x + \sqrt{V^2 - \epsilon^2}\sigma_z\bigr).
\end{equation}
Here, $\epsilon$ is the coherence parameter, and we fix $\epsilon=0.6$ for the experiments we run. The energy scale is denoted by $\hbar\Omega$, and $\sigma_x$ and $\sigma_z$ correspond to the standard Pauli matrices. We investigate protocols where the level-splitting parameter $V$ of the qubit and the temperature $T$ of the environment can be modulated.
We use the definitions outlined in \cite{brandner_thermodynamic_2020} to simulate and characterize the thermodynamics of the system and present it here again for convenience.

The generalized Bloch equations are 
\begin{equation}
    \begin{pmatrix} \dot{r}_x (t) \\ \dot{r}_y (t) \\ \dot{r}_z (t) \end{pmatrix} = 
    \begin{pmatrix} -k^+_{\lambdab_t} & -\Omega V_t & -\theta'_{V_t}\dot{V}_t 
                    \\ \Omega V_t & -k^+_{\lambdab_t} & 0
                    \\ \theta'_{V_t}\dot{V}_t & 0 & -2k^+_{\lambdab_t}
    \end{pmatrix} 
    \begin{pmatrix} r_x (t)\\ r_y (t)\\ r_z (t)\end{pmatrix}
    + \begin{pmatrix} 0 \\ 0 \\ k^-_{\lambdab_t} \end{pmatrix},
\end{equation}  
where 
\begin{equation}
    k^{\pm}_{\lambdab_t} \equiv \Gamma\Omega V \frac{1 \pm \exp{\hbar\Omega V/T}}{\exp[\hbar\Omega V/T] - 1}
\end{equation}
and
\begin{equation}
    \theta'_V \equiv \frac{\epsilon}{V\sqrt{V^2 - \epsilon^2}}.
\end{equation}
We compute the heat produced by the engine as
\begin{equation}
    Q = \frac{1}{2} \int_0^{\tau} \bigl(4 r_t  \arctanh (2r_t) + \ln[1/4 - r_t^2]\bigr)\ \dot{T}_t \ dt
\end{equation}
and the work done by the system as 
\begin{equation}
    W = -\hbar\Omega\int_0^{\tau} \bigl(-V_t\theta'_{V_t}r_t^x + r_t^z\bigr)\dot{V}_t \ dt,
\end{equation}
where the length of the Bloch vector 
$r_t \equiv \sqrt{(r_t^x)^2 + (r_t^y)^2 + (r_t^z)^2}$.
With these definitions, we compute the dissipation as
\begin{equation}
    \beta Q = \frac{1}{2} \int_0^{\tau} \frac{1}{T_t} \bigl(4 r_t  \arctanh (2r_t) + \ln[1/4 - r_t^2]\bigr)\ \dot{T}_t \ dt.
\end{equation}
Finally, as in \cite{brandner_thermodynamic_2020} we define the efficiency, 
\begin{equation}
    \eta =  \frac{W}{Q}.
    \label{eq:efficiency}
\end{equation}

From the Bloch vector, we can compute the density matrix $\rho_t$ as 
\begin{equation}
    \rho_t \equiv \mathbb{I}/2 + r_t^x \Pi^x_{V_t} + r_t^y \Pi^y_{V_t}  + r_t^z \Pi^z_{V_t},
\end{equation}
where 
\begin{align*} 
\Pi^x_{V_t} &= \lvert E_V^+ \rangle\langle E_V^- \rvert +  \lvert E_V^- \rangle\langle E_V^+ \rvert, \\ 
\Pi^y_{V_t} &= i \lvert E_V^-\rangle\langle E_V^+ \rvert - i\lvert E_V^+ \rangle\langle E_V^- \rvert, \\ 
\Pi^z_{V_t} &= \lvert E_V^+ \rangle\langle E_V^+ \rvert -  \lvert E_V^- \rangle\langle E_V^- \rvert.
\end{align*}
Lastly, we define 
\begin{align*}
    E_V^+ &=  \begin{pmatrix} \phantom{-} \sin[\theta_V/2] \\ -\cos[\theta_V/2] \end{pmatrix}, \\
    E_V^- &=  \begin{pmatrix} \cos[\theta_V/2] \\ \sin[\theta_V/2] \end{pmatrix}, \\
\end{align*}
with $\sin[\theta_V/2] = \epsilon/\sqrt{2V^2 + 2V\sqrt{V^2 - \epsilon^2}}$.

We first consider a base protocol $\lambdab(\bar{t}) = [T_{\bar{t}}, V_{\bar{t}}]$, where $\bar{t} = t/\tau$, and $\tau$ is the duration of the protocol. As in \cite{brandner_thermodynamic_2020}, we define $\lambdab(\bar{t})$ to be sinusoidal,
\begin{equation}
    \lambdab(\bar{t}) = \hbar\Omega[(1 + \sin^2(\pi\Omega\bar{t}), 1 + \sin^2(\pi\Omega \bar{t} + \pi/4)].
\end{equation}
To compute the optimal speed function $\phi_*$ that minimizes \eqref{eq:stot}, we first simulate the qubit engine under $\lambdab$ in the slow-driving regime with $\tau_{\rm slow} = 50$. Using the periodic steady-state $\rho_t$, we compute an ``instantaneous length'' of the path
\begin{equation}
    dL_{t_i} = \sqrt{\bigl(\rho_{t_i+dt}(\lvert 1 \rangle) - \rho_{t_i}(\lvert 1 \rangle)\bigr)^2 + \bigl(\rho_{t_i+dt}(\lvert 0 \rangle)  -\rho_{t_i}(\lvert 0 \rangle)\bigr)^2},
\end{equation}
where $0\leq i\leq\tau_{\rm slow}/dt$.
We then compute the discretized time-derivative of the speed function
\begin{equation}
    \dot{\phi}_{*_i} = \biggl(\frac{dt}{\tau_{\rm slow}}\biggr)\frac{L}{dL_{t_i}},
    \label{eq:speedot}
\end{equation}
where $L = \sum_{i} dL_{t_i}$. 
This leads to an effective time $\bar{t}_i = \sum_{k = 0}^{i}\dot{\phi}_{*_k}$, where $\dot{\phi}_* (\bar{t}_i) = \dot{\phi}_{*_i}$  and $0 \leq \bar{t}_i \leq 1$. Finally, the continuous function $\dot{\phi}_*(\bar{t})$ can be recovered via linear interpolation and $\phi_*(\bar{t})$ by integration. 
Intuitively, when the instantaneous path length $dL_{t_i}$ is larger than the average path length $\bigl(dt/\tau_{\rm slow}\bigr)\mathcal{L}$ for a time step, the protocol moves slower than a constant speed protocol in order to minimize the dissipative cost; when the instantaneous path length is smaller than the average path length the protocol moves faster than a constant speed protocol.

\begin{figure}[ht]
    \centering
    \includegraphics[width=0.45\linewidth]{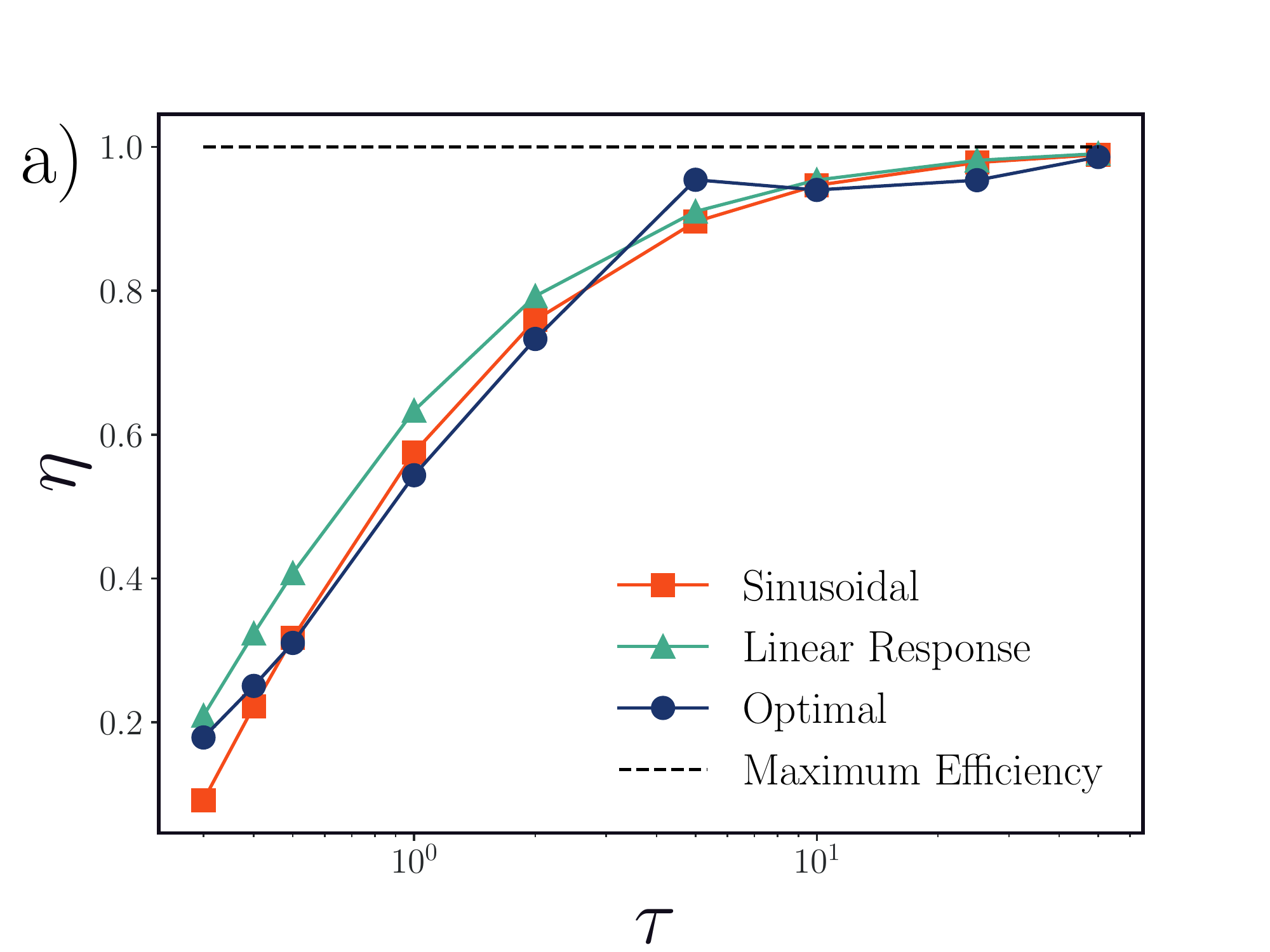}
    \includegraphics[width=0.45\linewidth]{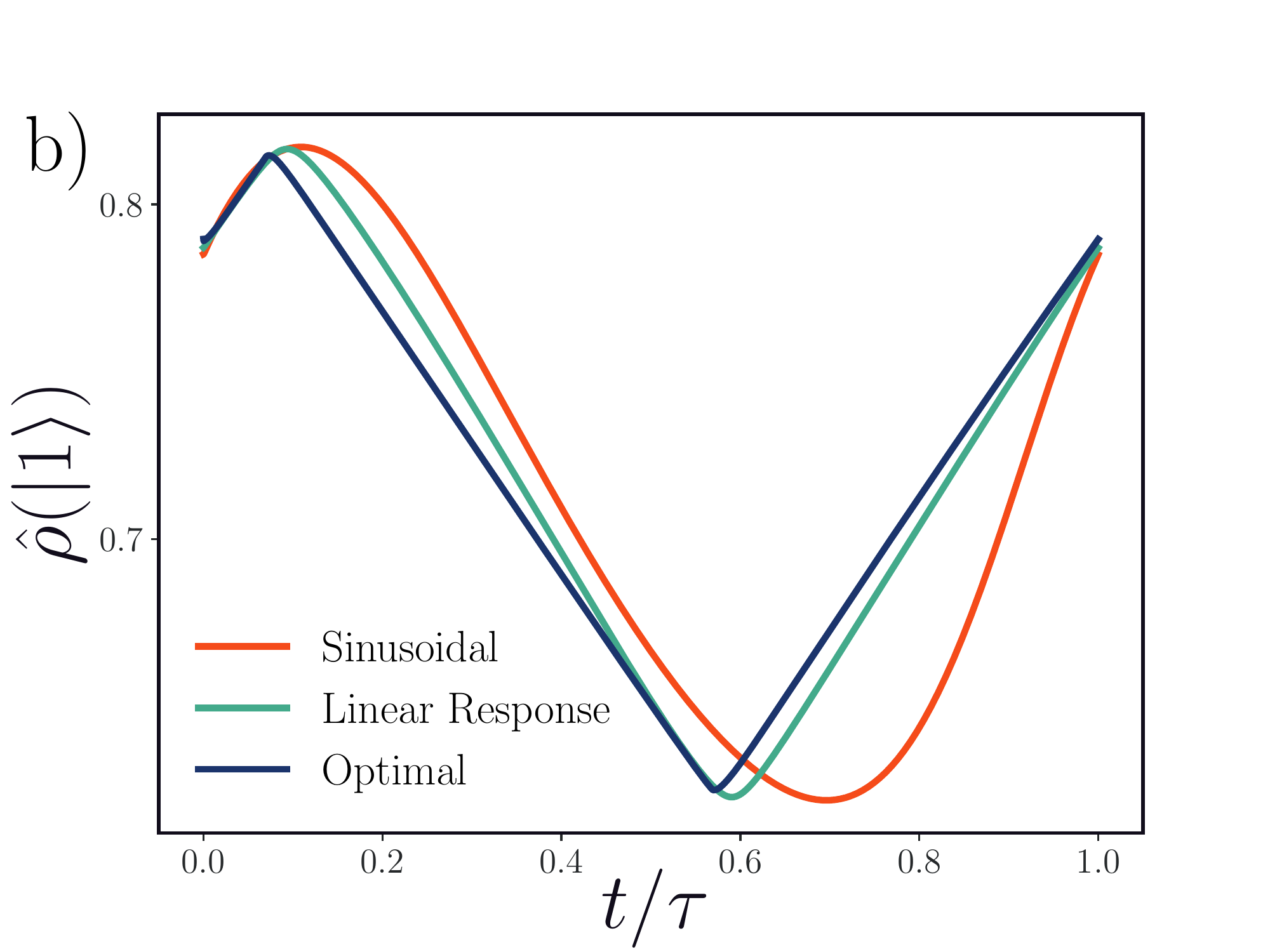}
    \caption{The efficiency (a) and steady-state densities at $\tau = 5$ (b) of the qubit engine for various speed functions for a prespecified sinusoidal protocol. The efficiency of the engine is higher under a speed function determined via a linear response approach (green triangle) compared to the speed function determined by minimizing \eqref{eq:stot} (dark blue circle). However, there is no guarantee of the intermediate states that the system visits when learning speed functions for a prespecified protocol.}
    \label{fig:qubit_dissip_eff_speed}
\end{figure}

We computed $\phi_{\rm LR} (t)$ using the linear response approximation presented in \cite{brandner_thermodynamic_2020}. In practice, the procedure for computing $\phi_{\rm LR} (t)$ is the same as computing $\phi_{*} (t)$, with a different definition of the length metric $\mathcal{L}$ (cf. Eq. S45 in \cite{brandner_thermodynamic_2020}). Fig.~\ref{fig:models} b) shows a comparison of the dissipation profiles of the qubit engine driven by $\lambdab(\phi_{*} (t))$, $\lambdab(\phi_{\rm LR} (t))$, and $\lambdab(t)$. We also plot the efficiency $\eta$ of the qubit engine driven under these protocols and observe that the efficiency remains the highest under $\lambdab(\phi_{\rm LR} (t))$ (Fig.~\ref{fig:qubit_dissip_eff_speed}). Optimizing \eqref{eq:stot} serves to minimize the dissipation, and there is no guarantee of a higher efficiency. Finally, we plot the steady-state density of the qubit engine driven by the three protocols, as shown in Fig.~\ref{fig:qubit_dissip_eff_speed}.

As evident in Fig~\ref{fig:qubit_dissip_eff_speed} b), there is no guarantee of the intermediate states that a system will visit when determining a speed function for a prespecified protocol $\lambdab$. Next, we instead consider learning a $\lambdab_*(t)$ using Algorithm~\ref{alg:autodiff} that cyclically drives the system through a set of prespecified intermediate distribution. We specified four intermediate distributions as the steady-state distributions the qubit engine driven under $\lambdab(t)$ at $t = 0, \tau/4, \tau/2,$ and $3\tau/4$ in the quasistatic limit. We then computed four intermediate displacement interpolations that interpolated these endpoints, where
\begin{equation}
    \rho (\lvert 1 \rangle, t) = \biggl(\frac{t_f - t}{t_f} \rho (\lvert 1 \rangle, 0) + \frac{t}{t_f} \rho (\lvert 1 \rangle, t_f)\biggr),
\end{equation}
$0\leq t\leq t_f$ and $\rho (\lvert 0 \rangle, t) = 1 -  \rho (\lvert 1 \rangle, t)$. 
Finally, we carried out the automatic differentiation optimization presented in Algorithm \ref{alg:autodiff} using a Mean Squared Error (MSE) objective
\begin{equation}
    \mathfrak{L}(t)= \frac{1}{2} \bigl(\rho_* (\lvert 0 \rangle, t) - \hat{\rho} (\lvert 0 \rangle,  t)\bigr)^2 + \bigl(\rho_* (\lvert 1 \rangle, t) - \hat{\rho} (\lvert 1 \rangle, t)\bigr)^2.
\end{equation}

\begin{figure}[ht]
    \centering
    \includegraphics[width=0.45\linewidth]{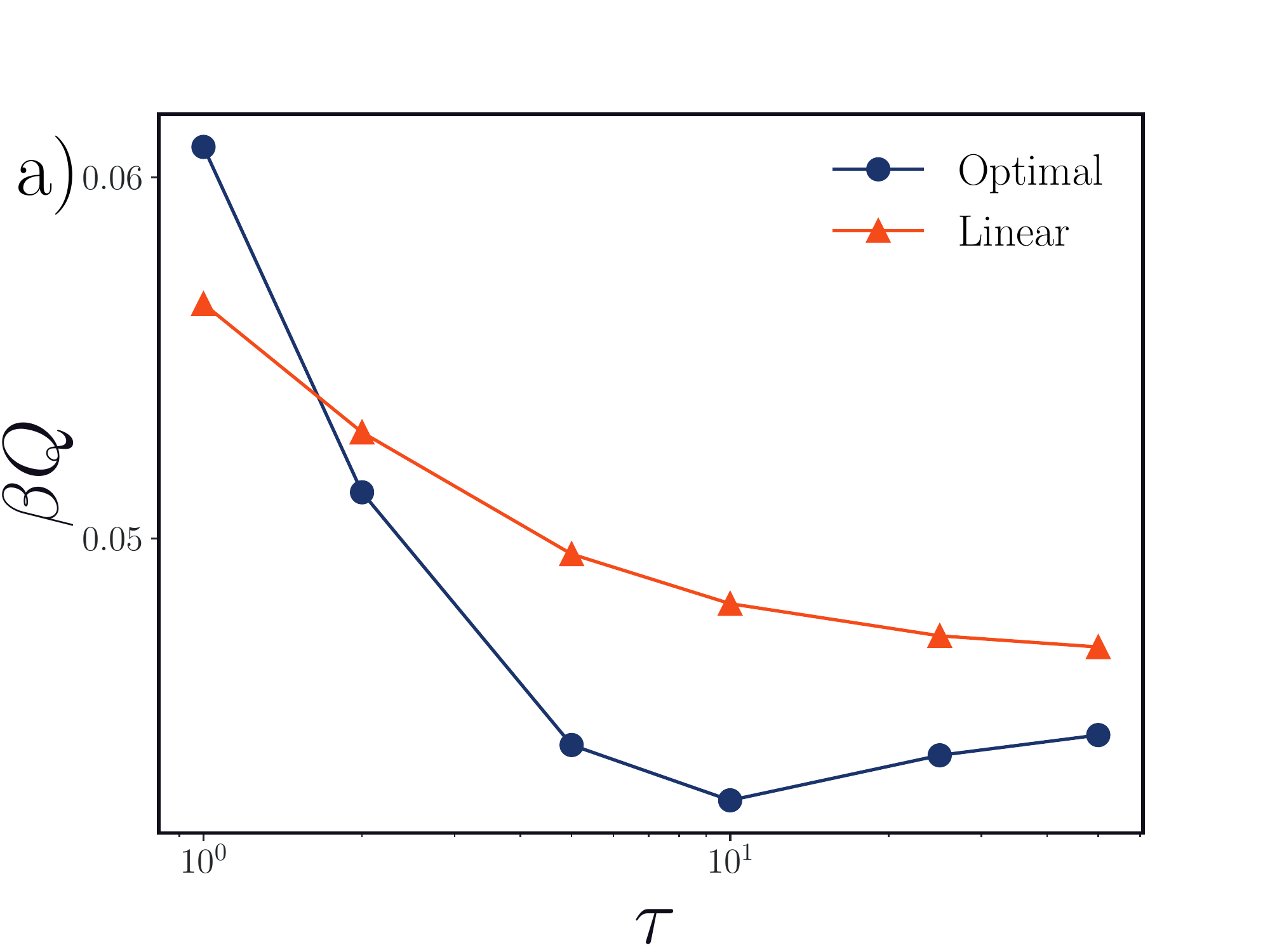}
    \includegraphics[width=0.45\linewidth]{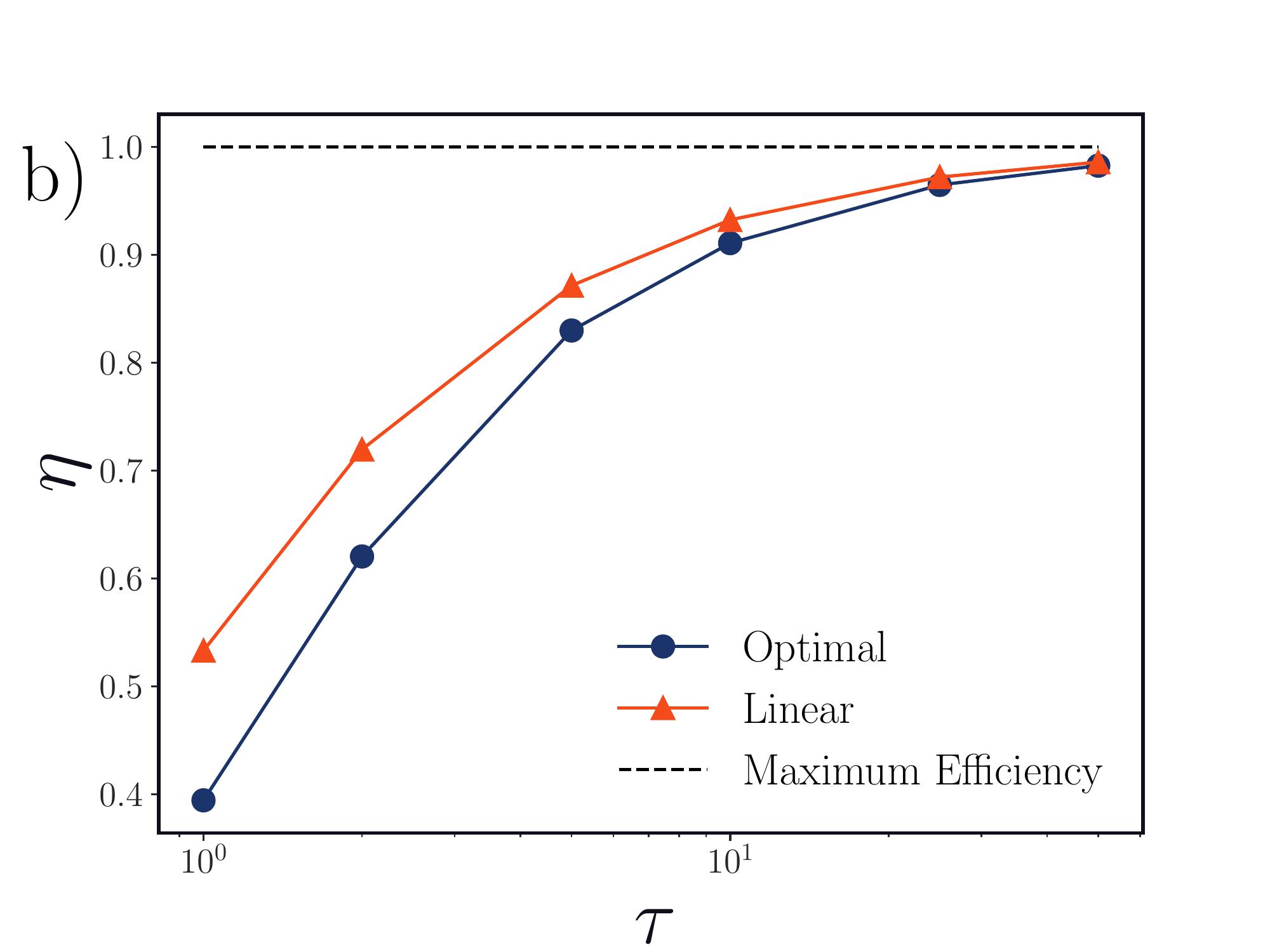}
    \includegraphics[width=0.45\linewidth]{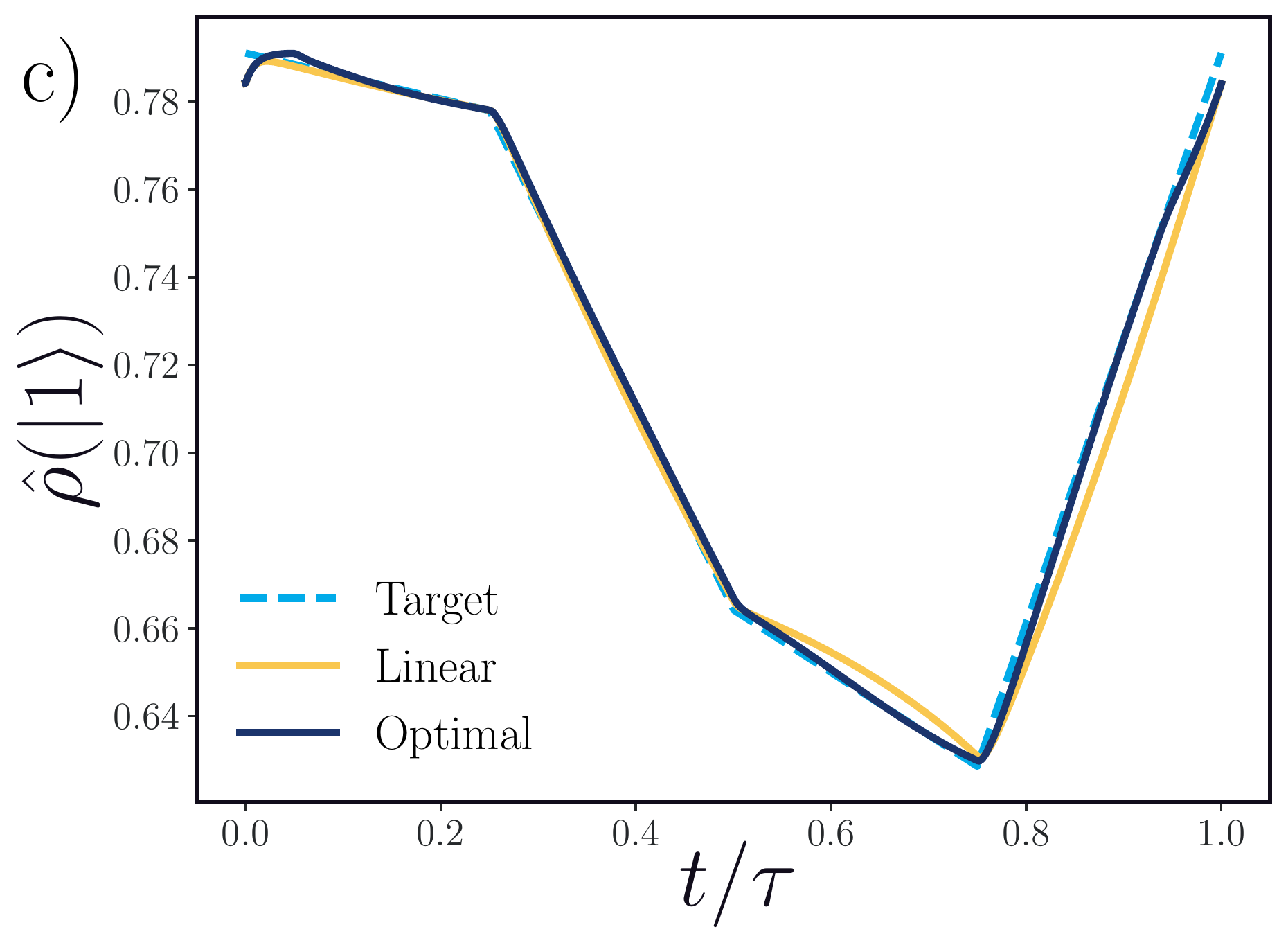}
    \caption{The dissipation (a), efficiency (b) and steady-state densities at $\tau=5$ (c) for protocols optimized to traverse through target intermediate distributions compared to a linear protocol. The dissipation is lower under the optimal protocol that minimizes \eqref{eq:stot} (dark blue circle) compared to a linear protocol (red triangle). For fastest-driving ($\tau = 1$), both protocols deviate significantly from intermediate target distributions, resulting in a high dissipative cost. For $\tau > 1$ optimal protocol achieves target intermediate distributions resulting in a lower dissipative cost. Still, the efficiency is higher for the  linear protocol compared to the optimal protocol. Finally, the optimal protocol closely realizes the target steady-state distribution, while the linear protocol slightly deviates from the target intermediate distributions for $t > \tau/2$ resulting in a higher dissipative cost.}
    \label{fig:qubit_int}
\end{figure}

For each of the four intervals of the protocol, $[0, \tau/4]$, $[\tau/4, \tau/2]$, $[\tau/2, 3\tau/4]$, and $[3\tau/4, \tau]$, we constrained the maximum and minimum $T$ and $V$ for $\lambdab_*$ to be the same as the maximum and minimum $T$ and $V$ for the base sinusoidal protocol $\lambdab$. Finally, we define $\lambdab_{\rm linear}$, a protocol which linearly interpolates $\lambdab(0)$, $\lambdab(\tau/4)$, $\lambdab(\tau/2)$, $\lambdab(3\tau/4)$, and $\lambdab(\tau)$. We compare the dissipation and efficiencies of the qubit engine driven by $\lambdab_*$ and $\lambdab_{\rm linear}$ (see Fig~\ref{fig:qubit_int}). We observe that under $\lambdab_*$ the qubit engine is less dissipative, but also slightly less efficient. Importantly, the former observation is consistent with~\eqref{eq:stot}; minimizing ~\eqref{eq:stot} does not necessarily guarantee higher efficiency. 

We compare these two protocols because we observe that the protocols generally have similar intermediate states for $\tau > 1$ (see Fig.~\ref{fig:qubit_int}). For the fastest driving, $\tau = 1$, both protocols deviate significantly from the target distribution as it is infeasible to learn a protocol $\lambdab_*$ that can achieve the target distributions on this timescale. However, we see that under $\lambdab_{\rm linear}$, the steady-state distribution of the qubit engine deviates slightly from the target displacement interpolation especially when $t > \tau/2$. This ultimately results in higher dissipation when driven by $\lambdab_{\rm linear}$ compared to $\lambdab_{*}$.

%\clearpage
%\bibliography{refs_app}

\end{document}